# Still Shrouded in Mystery: The Photon in 1925

Richard A. Campos
*Department of Physics and Astronomy*
*Lehman College, the City University of New York*
*Bronx, New York, 10468-1589, USA*

(15 February 2004)

Abstract

We present a translation of Albert Einstein's Rio de Janeiro manuscript on light quanta. In it, Einstein evaluates the Bohr-Kramers-Slater refutation of light quanta, which was concurrently the subject of intense empirical scrutiny on two continents. Written shortly before Heisenberg's discovery of quantum mechanics, the manuscript likely represents Einstein's last published remark on the constitution of light in the historical period known as the old quantum theory. It crystallizes the fact that by 1925 the light-quantum concept was still insufficient, even as the corpuscular properties of light gained decisive empirical confirmation.

## 1. Introduction

As we approach the centennial of Einstein's introduction of light quanta [1], it is well worth reconsidering just how long it takes for some very good ideas to make their way into the scientific schema. The journey can be tortuous, and in the case of the photon [2], it is great fun to recollect the legendary remarks about Einstein's "reckless" hypothesis [3], which should not be "held against him" for having "missed the target" [4]. In the particular case of Einstein's photons, two decades would pass before the conspiracy of theory and experiment left little option but to incorporate them into physical theory. But this is far from the whole story, because the 1905 light-quantum could not explain the wave qualities of radiation well known long before Einstein's arrival on the scene. The purpose of this work is to draw further attention to this point by way of a unique manuscript recently uncovered from the Einstein archives.

We remember the history of quantum mechanics as an encyclopedia with two great volumes. The first, the "old" quantum theory, begins with Max Planck's introduction in December 1900 of the energy quantum for material oscillators and the fundamental constant now bearing his name [5]; the second, the modern quantum theory, begins with Werner Heisenberg's discovery of quantum mechanics in June 1925 [6]. Both of these great advances resulted from intensely concentrated intellectual efforts, in Planck's case an "act of desperation" to resolve an inconsistency borne out of measured hot-body spectra, and in Heisenberg's case an inspiration "from the fog" of remote Helgoland island to resolve the insufficiency of classical mechanics to describe atomic phenomena. For both discoveries, the accent rests on the application of quantum principles to matter.

Between these two landmarks, Einstein championed a quantum hypothesis for light against a scientific community fixated on the wave picture of classical physics. The turning point for acceptance of the light quantum arrived in the early 1920's with Arthur Compton's analysis of frequency shifts for X-ray scattering from matter, modeled after a collision between a photon and a nearly free electron [7]. The additional confirmation of spatial correlations between the scattered light and the recoil electrons in 1925 finally put to rest those theoretical programs, such as that of Bohr, Kramers and Slater, which rejected light quanta outright. But the view of a photon as an independent particle bearing energy and momentum could still not account for optical interference phenomena. It is at this critical junction that Einstein packed his bags for a lecture circuit of South America, focused almost entirely on the relativistic theories. There is one remaining document which shows that Einstein's longstanding preoccupation with the photon extended well into the South American tour.

The manuscript of interest, "Remarks on the present situation of the theory of light," was penned by Einstein on May 7, 1925 for the express occasion of his visit to the Brazilian Academy of Sciences [8]. The fact that Einstein wrote about photons while in Rio de Janeiro is not widely known, let alone the fact that he added Brazil to his itinerary during the 'travel years' [9]. The letter is translated here for the first time in its entirety. The document importantly encapsulates Einstein's thoughts about the constitution of light just prior to Heisenberg's discovery of quantum mechanics on June 15. Since Einstein sailed the 12[th] of May from Rio on a two-week return trip to Berlin, the Rio manuscript represents Einstein's very last written impression of the photon as the curtains closed on the old quantum theory. It is significant not only because Einstein is still promoting the 1905 light-quantum hypothesis in early 1925, but also because it crystallizes the insufficiency of the hypothesis at the very moment when it is fully vindicated. The fact that the most brilliant physicist of the last century exhausted his means to achieve wave-particle synthesis using the classical methods at his disposal only accentuates the great need that was filled by Heisenberg's revolutionary discovery.

We place the scientific content of the manuscript in context with a brief review of the photon as a concept in evolution from 1905 to 1925. The personal circumstances of Einstein's South American tour is then summarized ahead of the full translation. The post-1925 period is discussed in the final remarks.

# 2. A brief history of the photon

## A. From independent particles to field equations

Einstein's "reckless" hypothesis wrestles Planck's discrete quanta of energy away from confined material oscillators, assigning them a reality of their own within the electromagnetic field that fills the intervening spaces. The year was 1905, Einstein's reputation was on the rise, and the heuristic basis of the light-quantum hypothesis was carefully advertised for everyone to see the in the very title of his acclaimed paper [1]. Einstein is very clear about his initial vision of the photon : "monochromatic radiation of low density… behaves thermodynamically as though it consisted of a number of independent energy quanta", and going even further, "the energy of a light ray spreading out from a point source is not continuously distributed over an increasing space but consists of a finite number of energy quanta which are localized at points in space, which move without dividing, and which can only be produced and absorbed as complete units."

This radical view of the photon as a Newtonian particle with separable and independent reality did not go well with the majority of Einstein's contemporaries. The main reason, simply stated, is that it could not account for the interference properties of light — after all, they withstood a century of empirical support before Einstein ever arrived on the scene. Thomas Young, father of the classic double-slit experiment, explained the basis of optical coherence like this: "In order that the effects of the two portions of the light may thus be combined, it is necessary that they be derived from the same origin, and that they arrive at the same point by different paths, in directions not much deviating from each other." [10] Different parts of a wave need to maintain a definite phase relationship.

There is plenty of evidence that this was the essential sticking point against the photon.  Important among the critics was Hendrik Lorentz, not only for his synergistic rapport with Einstein, but also because he nailed the case shut by incisive appeal to the technology of his day.  The argument he voiced at a conference in Utrecht on 17 April 1909 is this [11] : based on the latest world's records for interference (2 million ultraviolet fringes) and diffraction (150-cm Mount Wilson telescope), he figured that the photon extends by at least one meter or so in both the longitudinal and transverse directions.  How is a photon supposed to cohere like that, if it is localized and indivisible ?  The interference argument shows up in the Nobel lectures by Max Planck (1918) [12], Niels Bohr (1922) [13] and Robert Millikan (1924) [14].  It is carried over into the controversial Bohr-Kramers-Slater paper that concerns the Rio de Janeiro manuscript [15]. And, it is mentioned by Wolfgang Pauli and Walter Heitler, in their later recollections [16].  Evidently, the point really stuck, but we are jumping ahead a bit.

Whether stirred by Lorentz or not, we know that Einstein had already reached the same conclusion before January 1909,  because in that month he publishes an indirect proof of it using the old quantum theory [17].  The idea was to look at energy fluctuations out of the Planck hot-body spectrum, and when the derivation was done, Einstein sees the coexistence of  separate terms rooted on particle *and* wave phenomenologies.  No doubt about it, Einstein's photon had to have an extended property as well.  An important letter from Einstein to Lorentz dated 23 May 1909 confides [18]: "I am not at all of the opinion that one should think of light as being composed of mutually independent quanta localized in relatively small spaces."   The photon's burden of proof had just gotten larger, and it was up to Einstein to resolve it, after all he started the whole thing !

By early 1909, then, it was clear to Einstein that simple statistical arguments would have to give way to a proper wave equation of photons: "the fluctuation properties of radiation… offer few formal clues on which to build a theory."   In his May letter to Lorentz, he also presents the line of attack: look for localized solutions of nonlinear differential wave equations.   He explains the reason clearly in his autobiographical notes: "If the field-law is linear (and homogeneous), then the sum of two solutions is again a solution… it is impossible to deduce… an interaction between bodies, which can be described separately by means of solutions of the system. For this reason all theories up to now required, in addition to the field equations, special equations for the motion of material bodies under the influence of the fields" [19].

Einstein knew that a successful theory would have to account for both localized and extended properties.  For him, the familiar electric field of an electron serves a good starting analogy, and he motivates it by a simple dimensional argument linking Planck's constant to the electric charge, or, as Einstein calls it, the 'quantum of electricity' [20,21]. His optimism is plainly evident at the September 1909 conference of German scientists in Salzburg: "It probably wouldn't be as difficult to incorporate the interference phenomena as one thinks… I picture a quantum as a singularity surrounded by a large vector field. By using a large number of quanta one can construct a vector field that does not differ much from the kind of vector field we assume to be involved in radiations… I do not see any fundamental difficulty in the interference phenomena" [22].

Einstein is well rewarded by his optimism when he applies such ideas to the foundations of gravity during the decade of 1910. His intense efforts culminate in the spectacular confirmation of gravitational bending during a solar eclipse in 1919 [23], an event which transforms Einstein overnight into a recognizable and admired world figure [24]. But on the matter of photons, his program is far less successful. In 1924, he writes to Max Born: "my attempts to give tangible form to the quanta have foundered again and again" [25]. Part of the problem was the arrival of a remarkable new result to drive physics even farther from classical forms.

## B. The Bohr atom and the BKS controversy

Niels Bohr's theory of the quantized atom in 1913 [26] advances a successful account of emission line spectra from atomic elements. Bohr concludes that "energy radiation is not emitted (or absorbed) in the continuous way assumed in the ordinary electrodynamics, but only during the passing of the system between different 'stationary' states." Bohr's remarkable deduction is this: the fact that an atom emits or absorbs light at discrete frequencies predicates that its electronic *configurations* are also discrete. The frequency of light is no longer tied, as was in classical imagery, to the mechanical motion of orbiting electrons. This seems, as Planck emphasizes, a "monstrous" and "practically intolerable" demand. But numbers decide" [12].

So here was the photon, a bundle of energy very clear in hindsight, mediating the transitions between discrete atomic energy levels. Why did the debate on the reality of Einstein's light quanta not end here ? The answer is that the electromagnetic field can exist in vacuum, completely free of any atom, so it was still insufficient to also declare its energy structure discrete without additional evidence. Einstein used a good analogy here: "Even though beer is always sold in pint bottles, it does not follow that beer consists of indivisible pint portions." [27].

Niels Bohr tried to retain the classical electromagnetic field well into the 1920's. In his Nobel lecture [13], he argues as Lorentz did in 1909: "In spite of its heuristic value, however, the hypothesis of light-quanta, which is quite irreconcilable with so-called interference phenomena, is not able to throw light on the nature of radiation." But an asymmetry between quantized matter and nonquantized light raises disturbing questions. If light energy arrives in distributed fashion, but is absorbed only in bundles, what mechanism triggers an atomic jump between stationary states ? What establishes a correlation between atomic jumps in two atoms ?

Bohr tackles these questions with collaborators Hendrik Kramers and John Slater in January 1924 [15]. The BKS proposal assumes that atomic transitions are not externally triggered at all – an intrinsic 'virtual' field, oscillating at each of the Bohr transition frequencies,was supposed to determine them probabilistically. A consequence of the *independent* behavior of each atom is the abandon of "any attempt at a causal connexion between the transitions in distant atoms, and especially a direct application of the principles of conservation of energy and momentum, so characteristic for classical theories." The conservation theorems would hold on the average, but not for individual events. So, the price for ignoring the light quantum is very high indeed. Einstein had been frustrated in the past, but now he ponders giving up physics to become "a cobbler, or even an employee in a gaming-house." [25]. But before we review the important experiment of Walther Bose and Hans Geiger, which put BKS to the test, we need to also consider the final pillar of Einstein's light-quantum concept, the proposal of photon momentum.

## C. The photon carries momentum

While Einstein struggles with a constructive theory of photon interference, he moves brilliantly to sharpen the evidence for the localized behavior of photons. He does this by squeezing out yet another property – photon momentum – from statistical arguments for radiation pressure on a moving mirror [17]. In 1909, Einstein already declared: "*The elementary process of radiation seems to be directed*" (his italics) [20]. The great significance of this discovery is impressed by Wolfgang Pauli's opinion : "Einstein's report on the constitution of radiation at the physics meeting in Salzburg in 1909, where he appeared before a larger audience for the first time, can be considered as one of the landmarks of the development of theoretical physics. It … contains the important conclusion that the elementary process must be directed (needle radiation) not only for absorption, but also for emission of radiation, although this postulate was in open conflict with the classical idea of emission in a spherical wave" [16].

Einstein strengthens his conclusion in 1917 with a paper also considered a precursor to the discovery of the laser [28]. Instead of a mirror, he now imagines a moving gas molecule bathed by cavity radiation. Amazingly, Einstein finds that the Planck spectrum (and the Bohr jump condition, to boot) emerges only when the molecule experiences momentum transfer during absorption or emission processes. "If the molecule undergoes a loss of energy of magnitude h$\nu$ without external influence, by emitting this energy in the form of radiation (spontaneous emission), this process is also a *directed one*. There is no emission in spherical waves." The reality of the light quantum now seems inevitable to Einstein, but as he notes in 1918, "I still stand quite alone in this conviction." [29].

During the 1920's, the proving ground for photon momentum arrived not by way of research on visible light, but rather on X-rays [30]. The science of X-rays ( "As I will call the rays, for the sake of brevity" [31] ) was still young [32] when Einstein introduced the photon. At Salzburg in 1909, Johannes Stark rose up in support of the photon on the basis of X-rays : "I have in mind the phenomenon that even at great distances, up to 10 m, electromagnetic radiation that has left an X-ray tube for the surrounding space can still achieve concentrated action on a single electron" [22]. Does the extensive influence of the X-ray result from the packing of greater energy *and* momentum ?

Decisive evidence for photon momentum arrives in December of 1922, within days of Niels Bohr's Nobel lecture, as Arthur Compton reports the spectroscopic signature of scattered X-rays in matter [33]. In the classical view, electrons move in forced oscillations and so the scattered radiation is matched in frequency to the incident radiation. But something decisively nonclassical is going on here: many of the scattered X-rays have a lower frequency, and what is even more surprising, their shift does not depend on the incident frequency. The simplest model for this behavior, Compton [33] (and independently Peter Debye [34] ) suggest, is a billiard-like collision between an X-ray photon and a nearly free electron. The electron takes up some momentum when it is knocked off the atom, and this correlates to an outgoing X-ray of lower frequency. In Compton's words, "The corresponding effect with light would be for one's hand to appear green when illuminated with a blue light, to appear yellow when illuminated with green light, red when lighted by a yellow lamp and so on" [35].

The theory of the Compton effect afforded an extra benefit. Since it assumes energy-momentum conservation at the microscopic level, the formalism potentiates a key test of the BKS assumptions. What is needed is verification of the correlated exit *directions* of the recoiling electron and the scattered X-ray, using a suitable coincidence detector. A refutation of BKS is precisely what Bothe and Geiger in Berlin [36] , and also Compton with co-workers in Chicago [37], report within ten days of each other in April of 1925. From Bothe's Nobel lecture : "The readings of both counters were recorded side by side on a moving paper chart. In this way we succeeded after a few failures to establish the accuracy of any temporal 'coincidence' between the two pointer readings as being $10^{-4}$ sec. Film consumption however was so enormous that our laboratory with the film strips strung up for drying sometimes resembled an industrial laundry" [38]. Figure 1 reproduces a drawing from Compton's report to the National Academy of Sciences, which shows clearly the message borne out of the preliminary data from his research group [37].

The importance of BKS rests on the sharply antipodal position it took relative to the light quantum. With its demise, Einstein's 20-yr struggle to evince the corpuscular nature of light came to a successful end. But Einstein was not celebrating any victories at his Berlin residence. He was rather in South America on a solo three-month invited tour of Argentina, Uruguay, and Brazil, the backdrop to the Rio de Janeiro manuscript.

## 3. Einstein in South America

### A. Summary of the tour

Einstein's tour of South America, stretching from March through May 1925, occurs between his highly popularized visits to the United States (1921,1931), France (1922), Japan and Spain (1922-23) [39]. Out of his 'travel years,' the South American tour seems the forgotten middle child, receiving only minimalist exposure in the traditional literature. In one case even the basic chronology is inaccurately stated [9]. Thanks to the work of South American and other historians, however, it is now drawing greater interest [40-42], facilitated also by the online availability of Einstein's travel diary through the Einstein Archives in Jerusalem [43]. The political and scientific impact of Einstein's visit was extensive – ranging from amplified Zionism in the Jewish communities to overturned positivist attitudes of Brazilian physicists [42].

The story of how Einstein ends up in Rio de Janeiro begins in 1922 with membership to the League of Nations' International Committee on Intellectual Cooperation (ICIC), an invited position also shared by Marie Curie and Hendrik Lorentz (UNESCO, the United Nations Educational, Scientific and Cultural

Organization, is the modern version of the original directive). Through ICIC, Einstein meets Leopoldo Lugones, the Argentine journalist and writer who seminated the idea of a lecture tour in his country. By the time the invitation is accepted in 1924, Brazil and Uruguay are also added to the itinerary and Einstein prepares for a long absence from Berlin.

Looking into Einstein's diary [43], we see a physically demanding journey. He sails from Hamburg on March 5, 1925, making stops in Boulogne, Lisbon, Tenerife. He travels alone, and on March 14, he is warmed by a postcard for his 46th birthday. There are brief ports of call in Rio de Janeiro and Montevideo on March 21 and 24, on the way to Buenos Aires. In Argentina, a month-long circuit of lectures and receptions begins in the capital, then moves to La Plata and Córdoba. The last two weeks of the tour are divided between Montevideo (April 24-May1) and Rio de Janeiro (May 4-12), with a similarly packed agenda. There are some refreshing breaks. He is delighted by Rio's natural splendor ("An indescribable wealth of impressions in a few hours"), and by Montevideo's unpretentious elegance ("I found here love for one's country without any delusion of grandeur"). He has a few chances to practice his beloved violin [44]. Still, by anyone's measure, Einstein is justifiably exhausted. On May 3 he declares: "My nerves are spent," and on the last day of the tour he breathes: "Free at last, but more dead than alive." He is back in Berlin on June 1, 1925 [45].

Einstein's lectures in South America are almost exclusively dedicated to the relativistic theories. There is particular resonance here with the fact that the National Observatories of Rio de Janeiro and Córdoba played leading roles in the confirmation of Einstein's predictions for gravitational bending of light [46]. Their directors, Henrique Morize [47] and Charles Perrine, are by 1925 veteran chasers of gravitational bending, as they guided two solar eclipse expeditions on Brazilian soil (Passa Quatro in 1912 and Sobral in 1919). The theory of relativity is also known in the academic circles, consider for instance, that Rio de Janeiro hosted a lecture on that subject by the mathematician Émile Borel in 1922 [48]. Einstein is well aware of the South American participation in the landmark 1919 event and points out, "The problem conceived by my head, the luminous sky of Brazil had to solve" [42]. One can imagine the pride and honor that Morize experiences as Einstein enters Rio de Janeiro's National Observatory, and the Brazilian Academy of Sciences (BAS).

Without a doubt the visit to the BAS on May 7 is the most prestigious scientific event of Einstein's sojourn in Brazil. For this important occasion, the Academy names Einstein its first corresponding member and institutes a National Prize in his name. In the welcoming speech, Einstein is lauded for his work on relativity: "You brought a fundamental contribution to almost every field of physics. Without mentioning all your works, the study of Brownian motion, of specific heat, of the law of light-quanta, alone would suffice to consecrate you as an eminent physicist. However it was in the theory of relativity… that your genius was manifested" [49]. But Einstein was surely spent on relativity, because for this occasion he decides to discuss the reality of photons, despite the implied anticipations of the Academy ! The incident is unique – because no similar document appears from Einstein's other travels e.g. to the United States, Japan, Spain, France, where relativity was once again the focus of attentions.

## B. The Rio de Janeiro manuscript

A few words about how this letter appears in the Einstein Archives are of interest. Dr. Arthur Getúlio das Neves, head of Einstein's reception committee, filed the manuscript after the BAS lecture. It was translated into Portuguese and published in the inaugural issue of the BAS proceedings [8]. Neves died in 1928, and the letter was forgotten. In the 1990's, his grandson Jorge Getúlio Veiga reopened the stacks and placed a copy of it in the Einstein Archives. The present work provides a previously unavailable, complete, English translation.

Einstein's letter to the BAS is a concise statement about the photon concept as it stands in 1925, at the moment when the Bohr-Kramers-Slater proposal receives decisive empirical scrutiny in Chicago and Berlin. Einstein speaks about the photon as "controversial," with the tone of a provisional concept still subject to verification, and as yet unable to achieve a "logical synthesis" of continuous and localized optical properties. For Einstein, the vindication of the corpuscular light-quantum represented only a partial success. A full victory would only be savored if the light-quantum also appeared as part of a full theory of optical coherence. Evidently in 1925, the photon was still cloaked in mystery.

**Remarks on the Present Situation of the Theory of Light**

Until quite recently it was believed that, with the wave theory of light in its electromagnetic form, a conclusive knowledge was gained about the nature of radiation. However it is known for about 25 years that while this theory describes in a more precise way the geometric properties of light (refraction, diffraction, interference, etc), it cannot explain the thermal and energetic properties of radiation. A new theoretical concept, the quantum theory of light, which is close to Newton's emission theory, appeared suddenly beside the wave theory of light and gained through its achievements (explanation of Planck's radiation formula, of photochemical phenomena, Bohr's atomic theory) a sure position in science. Despite all the effort of physicists, a logical synthesis of the quantum and wave theories was unsuccessful to date. For this reason, the question about the reality of corpuscular light quanta is very controversial.

Recently N. Bohr, together with Cramers and Slater, conducted an interesting research which tried to theoretically explain the energetic properties of light, without invoking the hypothesis that the radiation consists of corpuscle-like quanta. According to the insight of these researchers, one should continue to envision that radiation consists of waves that propagate in all directions, and in the sense of the wave theory are steadily absorbed by matter, but nevertheless produce quantumlike effects in individual atoms following purely statistical laws, exactly as if radiation consisted of quanta of energy $h\nu$ and momentum $h\nu/c$. For the sake of this concept, the authors abandoned the exact validity of energy – and momentum – conservation theorems and in their place substituted a relation which only claims statistical validity.

To test this view empirically, the Berlin physicists Geiger and Bothe undertook an interesting experiment to which I would like to draw your attention. Some years ago, Compton derived a very important consequence from the quantum theory of light, which he confirmed by experiment. When hard Röntgen rays are scattered by electrons which constitute the atoms, it is possible for the momentum (impact) of the incoming quantum to be sufficiently large to knock an electron out of the atomic shell. The necessary energy for that is removed from the quantum during the collision and expresses itself, according to the principles of the quantum theory, as a reduced frequency of the scattered radiation compared to that of the incident Röntgen rays. This phenomenon, which was surely verified qualitatively and quantitatively by experiment, is called the 'Compton effect.'

In order to understand the Compton effect by the theory of Bohr, Cramers and Slater, one must treat scattered radiation as a continuous process in which all atoms of the scattering substance participate, while the ejection of the electrons has the character of isolated events following only statistical laws. According to the theory of light quanta, the scattering of light must also possess the character of isolated events, and there must exist an emitted electron in a definite direction each time the scattering radiation produces a secondary effect in matter. According to the theory of light quanta, there is a statistical dependence between the scattered radiation in Compton's sense and the electron emission, a dependence which in the theoretical view of the said authors must not exist.

In order to verify how it behaves in reality, one must have an instrument capable of registering a single elementary process of absorption and a single emitted electron, respectively. This instrument is made possible by an electrified needle wherein a single captured electron generates, through the secondary formation of ions, a momentary measurable discharge. With two such properly disposed needles, Geiger and Bothe can successfully answer the important question of statistical dependence or independence of the secondary processes mentioned above.

At the time of my departure from Europe, the experiment was not yet completed. However, according to the results attained so far, such statistical dependence appears to exist. If this is confirmed, then there is a new important argument for the reality of light quanta.

Einstein, 7.V.25

## 4. Final remarks

Einstein's letter to the Brazilian Academy of Sciences uniquely demarcates pre and post modern quantum mechanics. It summarizes the intensification of the quantum paradox relative to the constitution of light and chronologically represents the last known document of Einstein's on the subject – and realistically, the last opportunity for Einstein to do so before Heisenberg's creative leap in Helgoland. At that time, a photon theory which captured not only energy-momentum but also optical coherence was not available, despite two decades of Einstein's intellectual fervor. The insufficiency of the purely corpuscular photon of 1905 is echoed by many of Einstein's contemporaries, and is recognized by Einstein himself in 1909, possibly even earlier. "Physicists," writes Abraham Pais, "are conservative revolutionaries, resisting innovation as long as possible and at all intellectual cost, but embracing it when the evidence is incontrovertible" [50]. Millikan puts it cogently when it comes to the photon: "It may be said then without hesitation that it is not merely the Einstein [photoelectric] equation which is having extraordinary success at the moment, but the Einstein conception [of the photon] as well. But until it can account for the facts of interference and the other effects which have seemed thus far to be irreconcilable with it, we must withhold our full assent" [14]. This remark was written in 1924.

We reviewed in this paper how Einstein's constructive program to build coherence out of independent photons landed nowhere, so providing Planck, Bohr and numerous other quantum innovators a key reason not to abandon the classical electromagnetic field.  Einstein's travel diary to South America explicitly shows that he is preoccupied with photon coherence in 1925 even during his transatlantic crossing to Rio de Janeiro: "Idea for explanation of coherence from radiation emitted in different directions. 18 March" [51].  The statement, casual as it was without further elaboration, nevertheless adds supporting evidence that the open problem of the photon was still very much alive in Einstein's mind at the time.

We also believe that Einstein felt renewed urgency to address the problem of photon coherence leading up to 1925 and beyond.  Only a few months before leaving Berlin for South America, Einstein completed a series of papers on the quantum statistics of a molecular gas [52].  These are the papers well known for their treatment of Bose-Einstein condensation and for the proposal of atomic diffraction, all influenced by Louis de Broglie's discovery of matter waves earlier in 1924 [53].  Einstein begins to see that the rules of quantum statistics are quite different than those of classical physics, so that the "distribution of molecules among the [phase space] cells is not treated as a statistically independent one," reflecting perhaps "a mutual influence of the molecules – for the time being of a quite mysterious kind."   The fact that material particles also exhibit interference effects akin to light waves raised coherence once again to the forefront in 1925.  Einstein continued on the quest to bridge optical coherence well into 1926 [54], even as Erwin Schrödinger published his famous quantum-mechanical wave equation [55].  But by 1925 Einstein had exhausted his means, and physics decisively turned in the direction of quantum mechanics.

Ironically, the price paid for the photon's final assurance remained unacceptable to Einstein for the rest of his life.  To Don Howard, the truly revolutionary feature of quantum mechanics is its "mathematically consistent and empirically correct way of denying the mutual independence of interacting quantum systems" [56].  It is the non-separable, entangled quality of an interacting system's Schrödinger wave function which draws Einstein's final rejection of the new physics, immortalized by his famous controversial paper with Boris Podolsky and Niels Rosen [57].  But from Paul Dirac's treatment of the quantum electromagnetic field in 1927 [58], to Roy Glauber's theory of quantum-optical coherence in 1963 [59], Einstein's brain child, the photon, grows on the other hand ever more indispensable to physics.

# Acknowledgments

Permission to publish the Einstein letter is granted by The Albert Einstein Archives, The Hebrew University of Jerusalem, Israel. Permission to reproduce Compton's figure is granted by the Washington University in St. Louis. It is a pleasure to thank Barbara Wolff of the Albert Einstein Archives (Jerusalem), Robert Schulmann, Diana K. Buchwald and Rudy Hirschmann of the Einstein Papers Project (Caltech), Alfredo Tiomno Tolmasquim of the Museu de Astronomia e Ciências Afins (MAST/Rio de Janeiro), David Kasmier of the National Academy of Sciences, and Carole Prietto of the Washington University in St. Louis for their help with reproduction permissions. I also thank Dario Vasconcellos Campos of the Brazilian Consulate in New York, and Alex Lvovsky of the University of Konstanz for invaluable assistance with the translations. Harold Metcalf and Wolfgang Ketterle generously provided me copies of Einstein's papers listed in Ref 52. Marcus Chown is acknowledged for valuable discussions. This work was supported by the National Science Foundation under Grant No. 403350001.

reprints of other Einstein articles published in South America, including "Non-Euclidean geometry and physics," Revista Matemática Hispano-Americana **1,** 72-76 (1926), and "Pan-Europa," *La Prensa* (24 March 1925).

[42] Thomas F. Glick, "Between science and zionism: Einstein in Brazil," *Episteme, Porto Alegre* , 101-120 (July-December 1999). http://www.ilea.ufrgs.br/episteme.

[43] Albert Einstein, Diary of trip to South America between March 5 1925 and May 11 1925, Einstein Archives 29-132.

[44] Samuel Liondas, "El primer médico judio graduado en la Universidad Mayor de la República," *Noticias* n.100 (Sindicato Médico del Uruguay, September 1999). Einstein declined official lodging in Montevideo and stayed with the family of Dr. José Rossemblatt, who also played the violin: "El Ing. Luis P. Ponce, le dio da bienvenida, en nombre de la Municipalidad, ofreciéndole alojamento en el Parque Hotel, pero Einstein, agradeciéndole la atención, tuvo que declinarla ya que previamente lo habian invitado a residir en la casa de la familia Rossemblatt, en la Avenida 18 de Julio. Allí, nuestro joven medico le sirvióde cicerone durante su estadía en nuestro país, y también tuvo ocasión de practicar el violin junto a su ilustre huésped, ya que esta sabio, tal como lo declaró en una conferencia de prensa 'como todo matemático, era un entusiasta por la música.' " http://www.smu.org.uy.

[45] Albert Einstein, Letter to M. Besso (5 June 1925)*,* Ref. 25, p.121.

[46] Jean Einsenstaedt and Antonio Augusto Passos Videira, "A prova cearence das teorias de Einstein," Ciência Hoje **20** (November 1995) Translated: "The proof of Einstein's theories, from Ceará" (Brazil).

[47] Henrique Charles Morize is an important figure in Brazilian science history. Besides his pivotal role in the 1919 solar eclipse as director of Rio's National Observatory, he founded the Brazilian Academy of Sciences, won the adoption of Greenwich time zones (1913), brought in X-ray technology, and still found time to co-establish the country's first radio station (1923).

[48] Alfredo Tiomno Tolmasquim, "Constituição e diferenciação do meio cientifico brasileiro no contexto da visita de Einstein em 1925," Estudos Interdisciplinarios de America Latina y el Caribe, **7** (July-December 1996).

[49] Roberto Vergara Cafarelli, "Einstein no Brasil," in *Einstein e o Brasil*, Ref. 41, p. 120.

[50] '*Subtle is the Lord'*, Ref. 9, p.384.
19

# Figures and Figure Captions

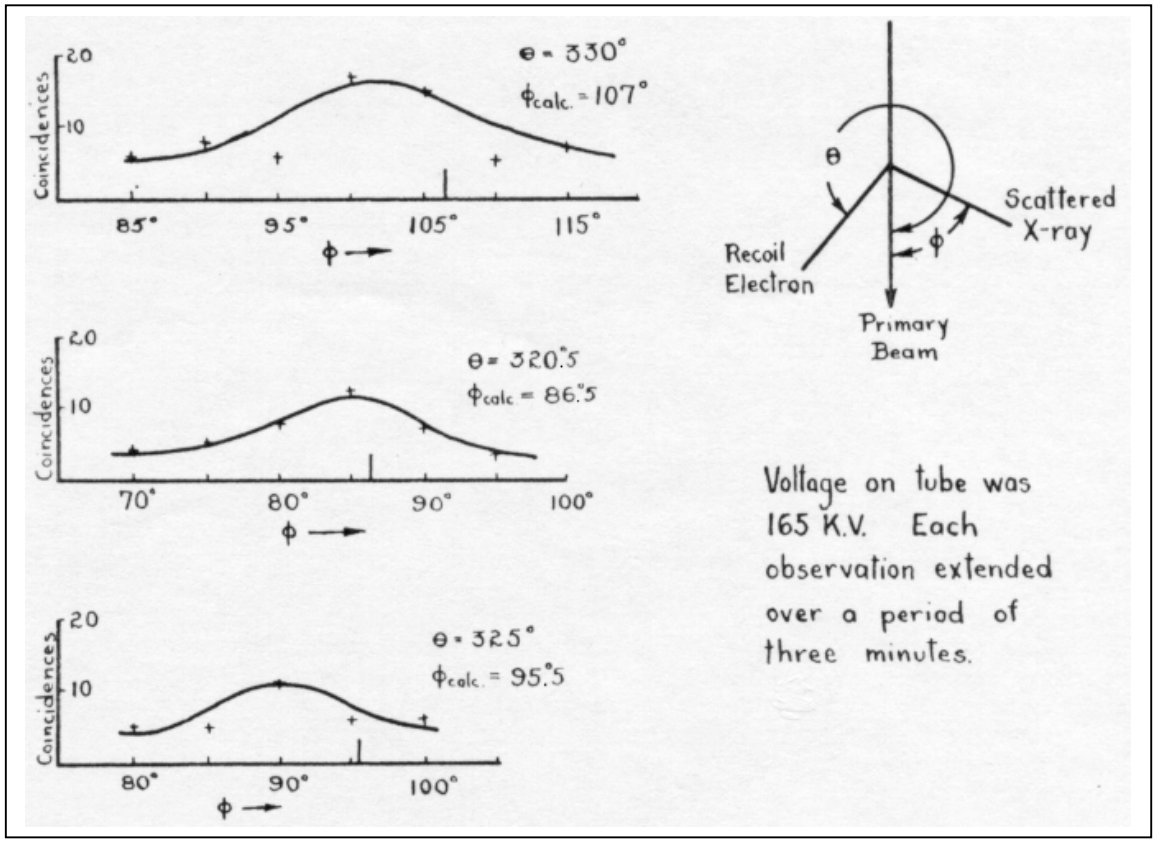

Figure 1. This reproduction from Arthur Compton's paper to the National Academy of Sciences shows the results of an auditory coincidence method, used by R. D. Bennett in his research group, to demonstrate simultaneity of a scattered photon and a recoil electron produced in the Compton effect. The correlated angular relationships more importantly serve to verify energy-momentum conservation. "Mr Bennett has connected each of a pair of head phones with one of the counting chambers through a 3-stage amplifier, and has listened for simultaneous impulses in the two phones. He uses a thin strip of mica to scatter the X-rays, and places the counting chambers inside an evacuated vessel to avoid effects due to the curving of recoil electrons in air. His results for three different settings of the electron counter are shown in Figure 1. The fact that coincidences are more frequent when the quantum counter is near the theoretical angle is in support of the quantum theory."

[Handwritten manuscript image]

Figure 2. Last paragraph of Einstein's Rio de Janeiro manuscript: "Zur Zeit meiner Abreise von Europa waren die Versuche noch nicht abgeschlossen. Nach den bisherigen Ergebnissen jedoch scheint statistiche Abhängigkeit vorzuliegen. Wenn sich dies bestätigt, so liegt ein neues wichtiges Argument für die Realität der Lichtquanten vor. A. Einstein, 7.V.25". Translation : "At the time of my departure from Europe, the experiment was not yet completed. However, according to the results attained so far, such statistical dependence appears to exist. If this is confirmed, then there is a new important argument for the reality of light quanta. A. Einstein, 7 May 1925." Twenty years passed before Einstein's 1905 light quanta attained decisive empirical verification. This was only a partial victory, because a quantum theory of light interference was not available at the time.